\documentclass{article}
\usepackage[margin=1.25in]{geometry}
\usepackage{graphicx}
\usepackage{amsmath}
\usepackage{amssymb}
\usepackage{mathtools}
\usepackage{parskip}
\usepackage[hidelinks]{hyperref}
\usepackage{graphicx}
\usepackage{natbib}
\usepackage[bf]{caption}
\usepackage{subcaption}

\usepackage{color}
\usepackage{amssymb}
\begin{document}
\title{
	On modelling bicycle power for velodromes: Part~I\\
	{\Large Formulation for individual pursuits}
}
\author{
	Michael A. Slawinski%
	\footnote{
		Memorial University of Newfoundland, Canada, \texttt{mslawins@mac.com}
	}\,,
	Rapha\"el A. Slawinski%
	\footnote{
		Mount Royal University, Canada, \texttt{rslawinski@mtroyal.ca}
	}\,,
	Theodore Stanoev%
	\footnote{
		Memorial University of Newfoundland, Canada, \texttt{theodore.stanoev@gmail.com}
	}
}
\date{September~3, 2020}
\maketitle
\begin{abstract}
For a moving bicycle, the power can be modelled as a response to the propulsion of the centre of mass of the bicycle-cyclist system.
On a velodrome, an accurate modelling of power requires a distinction between the trajectory of the wheels and the trajectory of the centre of mass.
We formulate and examine an individual-pursuit model that takes into account the aforementioned distinction.
In doing so, we provide details of the invoked physical principles and mathematical derivations, with an emphasis on their limitations.
We assume that a velodrome consists of two parallel straights and two semicircular arcs.
We neglect the effects of the track inclination along the straights and assume the track inclination along the curves to be constant.
For either segment, we consider two distinct black-line speeds.
For the latter, the lean-angle expression is derived based on a noninertial frame of the cyclist.
Among conclusions quantified by this model is the fact that a constant-cadence approach to an individual pursuit does not minimize the required power.
\end{abstract}
\section{Introduction}
In this article, we consider a mathematical model to account for power expenditure on velodromes.
This work is a mathematization of certain aspects of studies presented by \citet{MartinEtAl1998} and \citet{Underwood2012}.
The discussed model is pertinent to an individual pursuit of a cyclist following the black line in a constant aerodynamic position.
The circumference along the inner edge of this five-centimetre-wide line\,---\,also known as the measurement line and the datum line\,---\,corresponds to the official length of the track.

Using this model, for a given cyclist, we can calculate the power required to achieve a desired time or\,---\,since the relation between power and speed is one-to-one\,---\,the time achievable with a particular power.
Also, for repeated laps, we can estimate model parameters from the power and speed measurements.
The method proposed to estimate these parameters is specific to velodromes, and is different from the circuit study presented by \citet{Chung2012}.

We begin this article by presenting a model and justifying its mathematical formulation.
We illustrate both forward and inverse applications of the model.
We conclude by a discussion of results.
This article contains also three appendices.
\section{Formulation}
A mathematical model to account for the power required to propel a bicycle\,---\,along a straight course\,---\,with speed~$V_{\!\rightarrow}$ is \citep[e.g.,][]{DSSbici1}
\begin{align}
\label{eq:PV}
P&=F_{\!\leftarrow}\,V_{\!\rightarrow}\\
\nonumber&=\quad\frac{\overbrace{\,m\,g\sin\Theta\,}^\text{gravity}
+\!\!\!\overbrace{\,\,m\,a\,\,}^\text{change of speed}
+\overbrace{{\rm C_{rr}}\!\underbrace{\!\,m\,g\cos\Theta}_\text{normal force}
}^\text{rolling resistance}+\overbrace{\,\tfrac{1}{2}\,\eta\,{\rm C_{d}A}\,\rho\,(\!\!\underbrace{V_{\!\rightarrow}+w_{\leftarrow}}_\text{air flow speed}\!\!)^{2}\,}^\text{air resistance}}{\underbrace{\,1-\lambda\,\,}_\text{drivetrain efficiency}}\,V_{\!\rightarrow}\,,
\end{align}
where $F_{\!\leftarrow}$ stands for the forces opposing the motion and $V_{\!\rightarrow}$ for the ground speed.
In particular, $m$ is the mass of the cyclist and the bicycle, $g$ is the acceleration due to gravity, $\Theta$ is the slope of a hill, $a$ is the change of speed, $\rm C_{rr}$ is the rolling-resistance coefficient, $\rm C_{d}A$ is the air-resistance coefficient, $\rho$ is the air density, $w_{\leftarrow}$ is the wind component opposing the motion, $\lambda$ is the drivetrain-resistance coefficient, $\eta$ is a quantity that ensures the proper sign for the tailwind effect, $w_{\leftarrow}<-V_{\!\rightarrow}\iff\eta=-1$\,, otherwise, $\eta=1$\,.

To consider a steady ride,~$a=0$\,, on a flat course,~$\Theta=0$\,, in windless conditions,~$w=0$\,, we write
\begin{equation}
\label{eq:P}
P=\underbrace{\dfrac{{\rm C_{rr}}mg+\tfrac{1}{2}\,{\rm C_{d}A}\,\rho\,V_{\!\rightarrow}^2}{1-\lambda}}_{F_{\!\leftarrow}}V_{\!\rightarrow}\,.
\end{equation}

In modelling the power along a straight course, there is no distinction between the ground speed of the centre of mass and of any other point of the bicycle-cyclist system.
The distinction appears if the cyclist deviates from a straight course by leaning, which becomes pronounced on a velodrome.

Let us consider a velodrome, whose black-line distance is $S$\,, and the banking angle and radius are $\theta$ and $r$\,, respectively.
Also, let us assume that the centre of mass of the bicycle-cyclist system is $h$ above the ground, without lean.%
\footnote{We assume that the position of a cyclist on a bicycle remains the same, which is a reasonable assumption for an individual pursuit, after the initial acceleration.
Hence, $h$ is constant, and the change of the height of the centre of mass is due only to the lean angle,~$\vartheta$\,.}
If so, the radius of the centre-of-mass trajectory, $r_{\rm\scriptscriptstyle CoM}$\,, is shorter than $r$ by $h\sin\vartheta$\,, where $\vartheta$ is the angle, measured from the vertical, at which the cyclist leans.
Hence, the distance traveled\,---\,in one lap\,---\,by the centre of mass is shorter than the black line by
\begin{equation}
\label{eq:radii}
2\pi\,r-2\pi\,\overbrace{(r-h\sin\vartheta)}^{\displaystyle r_{\rm\scriptscriptstyle CoM}}=2\pi h\sin\vartheta\,.
\end{equation}
Thus\,---\,neglecting a progressive leaning and straightening between the straights and the circular arcs\,---\,the distance travelled by the centre of mass, along the straights and along the curves, for a single lap, is
\begin{equation}
\label{eq:DistLap}
S-2\pi r\qquad{\rm and}\qquad2\pi(r-h\sin\vartheta)\equiv2\pi\,r_{\rm\scriptscriptstyle CoM}\,,
\end{equation}
respectively.

\begin{figure}[h]
\centering
\includegraphics[scale=0.4]{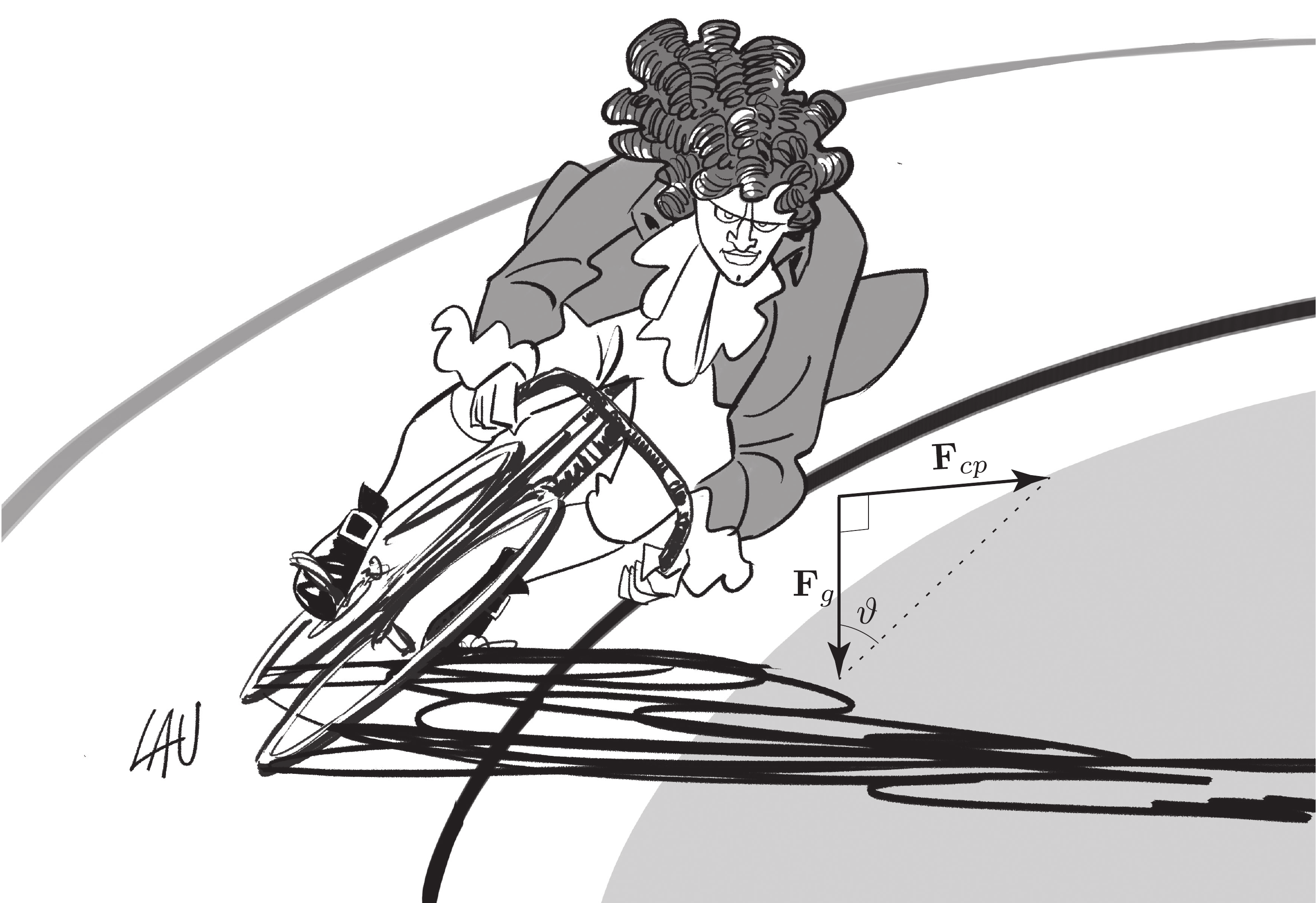}
\caption{\small Christiaan Huygens confident of a centripetal force, which he mathematized in~1659}
\label{fig:FigHuygens}
\end{figure}
Consider a laptime,
\begin{equation}
\label{eq:LapTime1}
t_\circlearrowleft=\dfrac{S-2\pi r}{\overline v_{\!\rightarrow}}+\dfrac{2\pi r}{\overline v_{\!\curvearrowleft}}\,,	
\end{equation}
where $v_{\!\rightarrow}$ and $v_{\!\curvearrowleft}$ are the black-line speed along the straights and the black-line speed along the curves, respectively; $\overline{\vphantom{V}\hphantom{v\,}}$ denotes the arithmetic average.
Also, the time along the curves is
\begin{equation}
\label{eq:LapTime2}
t_{\!\curvearrowleft}
=\dfrac{2\pi r}{\overline v_{\!\curvearrowleft}}
=\dfrac{2\pi\,r_{\rm\scriptscriptstyle CoM}}{\overline V_{\!\curvearrowleft}}\,,	
\end{equation}
where $\overline V_{\!\curvearrowleft}$ is the average centre-of-mass speed along the curves; $\overline v_\rightarrow\equiv\overline V_{\!\rightarrow}$\,, since\,---\,along the straights\,---\,there is no distinction between the speed of the centre of mass and the speed of any point of the bicycle-cyclist system.

The lean angle of a cyclist\,---\,as illustrated in Figures~\ref{fig:FigHuygens} and \ref{fig:FigFreeBody}, and entailed by expression~(\ref{eq:Fc_vartheta}), below, as well as discussed in Appendix~\ref{app:LeanAng}\,---\,is
\begin{equation}
\label{eq:vartheta}
\vartheta=\arctan\dfrac{F_{cp}}{F_g}\,,
\end{equation}
where the magnitude of the centripetal force is
\begin{equation}
\label{eq:Fc}
F_{cp}=\dfrac{m\,V_{\!\curvearrowleft}^{\,2}}{r_{\rm\scriptscriptstyle CoM}}\,,
\end{equation}
and of the force of gravity is~$F_g=m\,g$\,; hence,
\begin{equation}
\label{eq:Num3}
\overline V_{\!\curvearrowleft}=\sqrt{g\,r_{\rm\scriptscriptstyle CoM}\tan\vartheta}\,.
\end{equation}
Inserting expression~(\ref{eq:Num3}) into expression~(\ref{eq:LapTime2}), we obtain
\begin{equation}
\label{eq:CircArc}
\overline v_{\!\curvearrowleft}=r\sqrt{\dfrac{g\tan\vartheta}{r_{\rm\scriptscriptstyle CoM}}}\,.
\end{equation}
Combining expressions~(\ref{eq:LapTime1}), (\ref{eq:LapTime2}) and (\ref{eq:Num3}), we write
\begin{equation}
\label{eq:LapTime}
t_\circlearrowleft=\dfrac{S-2\pi r}{\overline V_{\!\rightarrow}}+2\pi\sqrt{\dfrac{r_{\rm\scriptscriptstyle CoM}}{g\tan\vartheta}}\,.	
\end{equation}

Given $v_\rightarrow$ and $v_{\!\curvearrowleft}$\,, with $r$\,, $S$ and $h$ assumed to be known, we use equations~(\ref{eq:LapTime1}) and (\ref{eq:CircArc}) to find $t_\circlearrowleft$ and $\vartheta$\,, respectively.
Then, we use expression~(\ref{eq:Num3}) to find $\overline V_{\!\curvearrowleft}$\,.

Following expressions~(\ref{eq:DistLap}), the proportion of distance travelled, per lap, by the centre of mass, is
\begin{equation*}
1-\dfrac{2\pi r}{S}\qquad{\rm and}\qquad\dfrac{2\pi(r-h\sin\vartheta)}{S}\,,
\end{equation*}
along the straights and the curves, respectively.
Hence, according to the harmonic average, discussed in Appendix~\ref{app:HarmAve}, the average centre-of-mass speed, per lap, is
\begin{subequations}
\label{eq:AveV}
\begin{align}
\langle V\rangle&=\dfrac{1}{\dfrac{1}{\overline V_{\!\rightarrow}}\left(\dfrac{1-\dfrac{2\pi r}{S}}{1-\dfrac{2\pi h\sin\vartheta}{S}}\right)+\dfrac{1}{\overline V_{\!\curvearrowleft}}\left(\dfrac{\dfrac{2\pi(r-h\sin\vartheta)}{S}}{1-\dfrac{2\pi h\sin\vartheta}{S}}\right)}\label{eq:AveVa}\\
\nonumber\\
&=\dfrac{\overline{V}_{\!\rightarrow}\overline{V}_{\!\curvearrowleft}\left(S-2\pi h\sin\vartheta\right)}{S\,\overline{V}_{\!\curvearrowleft}+2\pi\left(r\left(\overline{V}_{\!\rightarrow}-\overline{V}_{\!\curvearrowleft}\right)-\overline{V}_{\!\rightarrow}\,h\sin\vartheta\right)}\label{eq:AveVb}
\,.
\end{align}
\end{subequations}
The average power per lap is
\begin{subequations}
\label{eq:power}
\begin{align}
\overline P&=\dfrac{1}{1-\lambda}\,
\underbrace{\dfrac{\overline{V}_{\!\rightarrow}\overline{V}_{\!\curvearrowleft}\left(S-2\pi h\sin\vartheta\right)}{S\,\overline{V}_{\!\curvearrowleft}+2\pi\left(r\left(\overline{V}_{\!\rightarrow}-\overline{V}_{\!\curvearrowleft}\right)-\overline{V}_{\!\rightarrow}\,h\sin\vartheta\right)}}_{\langle V\rangle}\,\left\{\vphantom{\dfrac{2\pi\overbrace{(r-h\sin\vartheta)}^{r_{\rm CoM}}}{S}}\right.\label{eq:modelO}\\
&\left.\left.{\rm C_{rr}}\,mg\left(1-\dfrac{2\pi r}{S}\right)\right.\right.\label{eq:modelA}\\
+&\left.\left.\left({\rm C_{rr}}\underbrace{m\,g\,(\sin\theta\tan\vartheta+\cos\theta)}_N\cos\theta
+{\rm C_{sr}}\Bigg|\underbrace{m\,g\,\frac{\sin(\theta-\vartheta)}{\cos\vartheta}}_{F_f}\Bigg|\sin\theta\right)
\dfrac{2\pi\overbrace{(r-h\sin\vartheta)}^{\displaystyle r_{\rm\scriptscriptstyle CoM}}}{S}\right.\right.\label{eq:modelB}\\
+&\left.\tfrac{1}{2}\,{\rm C_{d}A}\,\rho\,\left(\left(1-\dfrac{2\pi r}{S}\right)\overline V_{\!\rightarrow}^{\,2}+\dfrac{2\pi\overbrace{(r-h\sin\vartheta)}^{\displaystyle r_{\rm\scriptscriptstyle CoM}}}{S}\overline V_{\!\curvearrowleft}^{\,2}\right)\right\}\label{eq:modelC}
\,.
\end{align}	
\end{subequations}
Herein, $\rm C_{sr}$ is the coefficient of the lateral friction and $\big|{\,\,}\big|$ stands for the magnitude.
If $S\rightarrow\infty$\,, expression~(\ref{eq:power}) reduces to expression~(\ref{eq:P}), as expected.

The second fraction in factor~(\ref{eq:modelO}) is the average centre-of-mass speed, per lap, which combines, proportionally, $\overline V_{\!\rightarrow}$\,, along the straights, and $\overline V_{\!\curvearrowleft}$\,, along the banks.
Summand~(\ref{eq:modelA}) is the rolling resistance along the straights.%
\footnote{\label{foot:Straight}A modern velodrome, such as V\'elodrome de Bordeaux-Lac, has $S=250\,{\rm m}$\,, $r=23\,{\rm m}$\,, $\theta=\pi/4\,{\rm rad}\equiv45^\circ$\,, along the banks, ${\rm\Theta}=0.23\,{\rm rad}\equiv13^\circ$\,, along the straights; the latter angle is not zero to avoid an excessive change of track inclination and to allow for a gentle slope transition between the banks and the straights.
We could refine our model by including
\begin{equation*}
\left({\rm C_{rr}}\cos^2\!{\rm\Theta}+\rm C_{sr}\,\sin^2\!{\rm\Theta}\right)\,m\,g\,\left(1-\dfrac{2\pi r}{S}\right)\,,
\end{equation*}
to account for the force of lateral friction along the straights, which would be analogous to summand~(\ref{eq:modelB}), with $\vartheta=0$\,.
However, since ${\rm\Theta}$ is small, $\cos{\rm\Theta}\approx1$ and $\sin{\rm\Theta}\approx0$\,, we choose to consider only
\begin{equation*}
{\rm C_{rr}}\,m\,g\,\left(1-\dfrac{2\pi r}{S}\right)\,,
\end{equation*}
which is summand~(\ref{eq:modelA}).}
Summand~(\ref{eq:modelC}) is the air resistance, which is a function of the proportions between $\overline V_{\!\rightarrow}^{\,2}$\,, along the straights, and $\overline V_{\!\curvearrowleft}^{\,2}$\,, along the banks.%
\footnote{\label{foot:Wheels}We could refine the model by including the effect of air resistance of rotating wheels \citep[Appendix~D]{DSSbici1}, which would require introducing another resistance coefficient to summand~(\ref{eq:modelC}), if the wheels are the same, or two coefficients, if they are different.}

\begin{figure}
\centering
\includegraphics[scale=0.7]{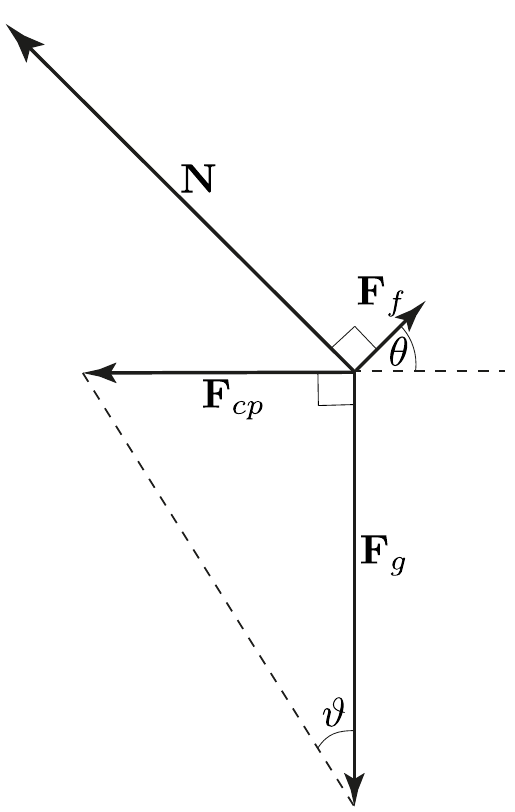}
\caption{\small Force diagram: Inertial frame}
\label{fig:FigFreeBody}
\end{figure}
To formulate summand~(\ref{eq:modelB}), we use the relations among the magnitudes of vectors~$\bf N$\,, ${\bf F}_g$\,, ${\bf F}_{cp}$ and ${\bf F}_f$\,, illustrated in Figure~\ref{fig:FigFreeBody}.
In accordance with Newton's second law, for a cyclist to maintain a horizontal trajectory, the resultant of all vertical forces must be zero,
\begin{equation}
\label{eq:Fy0}
\sum F_y=0=N\cos\theta+F_f\sin\theta-F_g
\,.
\end{equation}
In other words, ${\bf F}_g$ must be balanced by the sum of the vertical components of normal force, $\bf N$\,, and the friction force, ${\bf F}_f$\,, which is parallel to the velodrome surface and perpendicular to the instantaneous velocity.
Depending on the centre-of-mass speed and the radius of curvature for the centre-of-mass trajectory, if $\vartheta<\theta$\,, ${\bf F}_f$ points upwards, in Figure~\ref{fig:FigFreeBody}, which corresponds to its pointing outwards, on the velodrome;
 if $\vartheta>\theta$\,, it points downwards and inwards.
 If $\vartheta=\theta$\,, ${\bf F}_f=\bf 0$\,.
 Since we assume no lateral motion, ${\bf F}_f$ accounts for the force that prevents it.
 Heuristically, it can be conceptualized as the force exerted in a lateral deformation of the tires.

For a cyclist to follow the curved bank, the resultant of the horizontal forces,
\begin{equation}
\label{eq:Fx0}
\sum F_x=-N\sin\theta+F_f\cos\theta=-F_{cp}
\,,
\end{equation}
is the centripetal force,~${\bf F}_{cp}$\,, whose direction is perpendicular to the direction of motion and points towards the centre of the radius of curvature.
According to the rotational equilibrium about the centre of mass,
\begin{equation}
\label{eq:torque=0}
\sum\tau_z=0=F_f\,h\,\cos\left(\theta-\vartheta\right)-N\,h\,\sin(\theta-\vartheta)\,,
\end{equation}
where $\tau_z$ is the torque about the axis parallel to the instantaneous velocity, which implies
\begin{equation}
\label{eq:torque}
F_f=N\tan(\theta-\vartheta)\,.	
\end{equation}
Substituting expression~(\ref{eq:torque}) in expression~(\ref{eq:Fy0}), we obtain
\begin{equation}
\label{eq:N}
N=\dfrac{m\,g}{\cos\theta-\tan(\theta-\vartheta)\sin\theta}
=m\,g\,(\sin\theta\tan\vartheta+\cos\theta)
\,.
\end{equation}
Using this result in expression~(\ref{eq:torque}), we obtain
\begin{equation}
\label{eq:Ff}
F_f=m\,g\,(\sin\theta\tan\vartheta+\cos\theta)\tan(\theta-\vartheta)
=m\,g\,\dfrac{\sin(\theta-\vartheta)}{\cos\vartheta}\,.
\end{equation}
Since the lateral friction, ${\bf F}_f$\,, is a dissipative force, it does negative work.
Hence, the work done against it\,---\,as well as the power\,---\,needs to be positive.
For this reason, in expression~(\ref{eq:modelB}), we consider the magnitude of ${\bf F}_f$\,.

To relate $F_{cp}$ and $\vartheta$\,, we use results~(\ref{eq:N}) and (\ref{eq:Ff}) in expression~(\ref{eq:Fx0}), to obtain
\begin{equation}
\label{eq:Fc_vartheta}
F_{cp}=N\sin\theta-F_f\cos\theta=m\,g\,\tan\vartheta\,,
\end{equation}
which is tantamount to expression~(\ref{eq:vartheta}).
Examining expressions~(\ref{eq:Fc}) and (\ref{eq:Fc_vartheta}), we see that the lean angle is a function of the centre-of-mass speed and of the radius of curvature for the centre-of-mass trajectory; it is independent of mass or the track inclination.

\begin{figure}[h]
\centering
\includegraphics[scale=0.7]{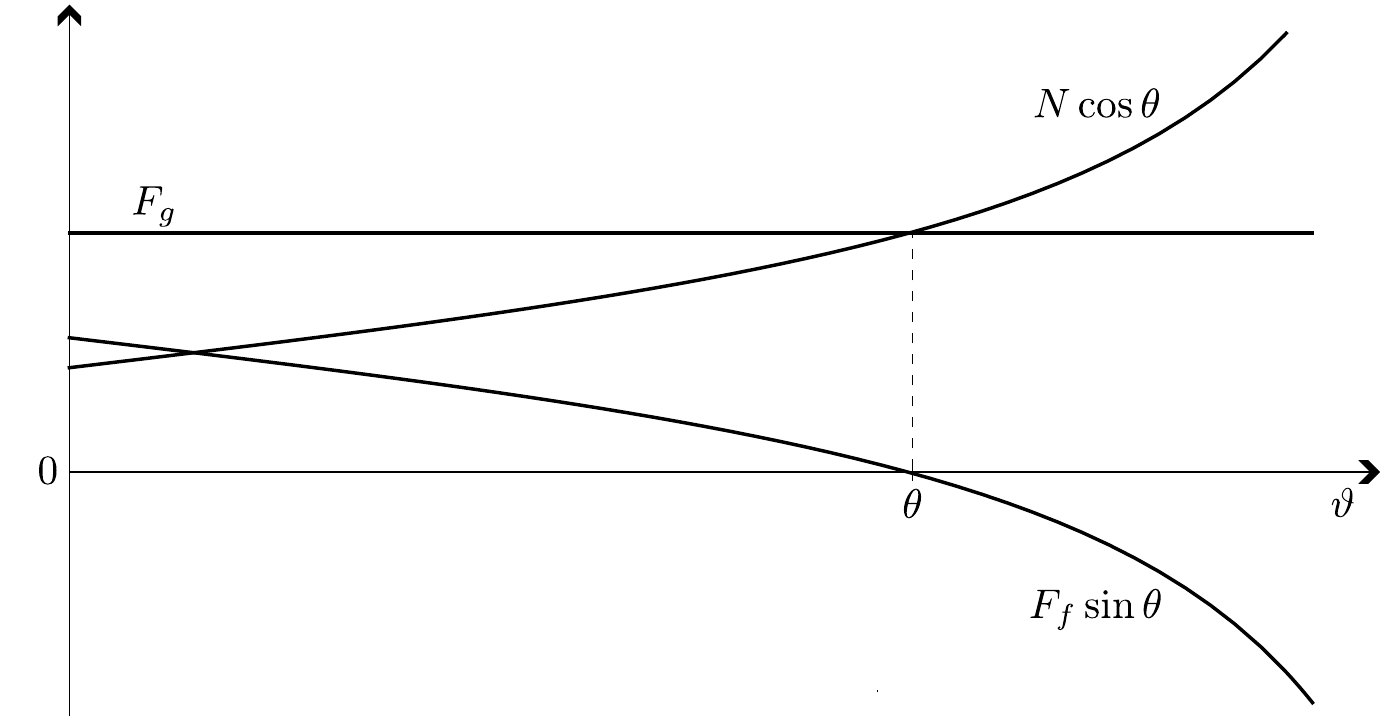}
\caption{\small $F_g$ in terms of vertical components of $\bf N$ and ${\bf F}_f$}
\label{fig:FigNFf}
\end{figure}
In terms of solutions~(\ref{eq:N}) and (\ref{eq:Ff}), expression~\eqref{eq:Fy0}\,---\,as a function of~$\vartheta$\,, for a fixed value of~$\theta$\,---\,is shown in Figure~\ref{fig:FigNFf}.
$F_g$~is constant, as required.
Also, as required, $F_f=0$ and $N\cos\theta=F_g$\,, at $\vartheta=\theta$\,.
For $\vartheta<\theta$\,, ${\bf F}_f$ points outwards, hence\,---\,in accordance with Figure~\ref{fig:FigFreeBody}\,---\,it is positive.
For $\vartheta>\theta$\,, ${\bf F}_f$ points inwards and, hence, is negative.
The crossing of two curves corresponds to $\vartheta$ at which the vertical components of $\bf N$ and ${\bf F}_f$ are equal to one another.
\section{Numerical example}
\label{sec:NumEx}
\subsection{Forward model}
\footnote{For consistency with power meters, whose measurements are expressed in watts,~$\rm W$\,, which are $\rm{kg\,m^2/s^3}$\,, we use the {\it SI} units for all quantities.
Mass is given in kilograms,~$\rm{kg}$\,, length in meters,~$\rm{m}$\,, and time in seconds,~$\rm{s}$\,; hence, area is in $\rm{m^2}$\,, speed is in $\rm{m/s}$ acceleration in $\rm{m/s^2}$ and force in $\rm kg\,m/s^2=:N$\,; angles are in radians, $\rm{rad}$\,, as well as in degrees,${}^\circ$\,.}%
Let us consider the following values.
For the bicycle-cyclist system, $m=111\,{\rm kg}$\,, $h=1.2\,{\rm m}$\,, ${\rm C_{d}A}=0.2\,{\rm m^2}$\,, ${\rm C_{rr}}=0.002$\,, ${\rm C_{sr}}=0.003$ and $\lambda=0.02$\,.
For the velodrome, $S=250\,{\rm m}$\,, $r=23\,{\rm m}$ and $\theta=0.7505\,{\rm rad}\equiv 43^\circ$\,.
For the external conditions, $g=9.81\,{\rm m/s^2}$ and $\rho=1.225\,{\rm kg/m^3}$\,.

Let the laptime be such that, according to expression~(\ref{eq:LapTime1}), the corresponding black-line speed is $v_\rightarrow=v_{\!\curvearrowleft}=\overline V_{\!\rightarrow}=12\,{\rm m/s}$\,; a constant black-line speed is tantamount to a constant cadence, which is a common approach for a workout or even for a pursuit.
In accordance with expressions~(\ref{eq:LapTime2}) and (\ref{eq:Num3}), $\vartheta=0.555468\,{\rm rad}\equiv 31.8260^\circ$\,, and $\overline V_{\!\curvearrowleft}=11.6698\,{\rm m/s}$\,.
Hence, in accordance with expression~(\ref{eq:power}), $\overline P=229.6723\,{\rm W}$\,.
Also, in accordance with expression~(\ref{eq:AveV}), the average centre-of-mass speed, per lap, is $\overline V_{\!\curvearrowleft}<\langle{V}\rangle=11.8091\,{\rm m/s}<\overline V_{\!\rightarrow}$\,, as expected.

The values of summands~(\ref{eq:modelA}),~(\ref{eq:modelB}) and (\ref{eq:modelC}), which represent the forces opposing the movement, are, respectively, $0.9189\,{\rm N}$\,, $1.3195\,{\rm N}$ and $16.8213\,{\rm N}$\,.
The air resistance, which is the third summand, has the dominant effect, with respect to the other summands, which are associated with the effect of the wheel contact with the velodrome surface.

If the centre of mass is not taken into account, expression~(\ref{eq:P}) results in $\overline P=242.6672\,{\rm W}$\,, which is an overestimate.
The values of the first and second summands in the numerator of expression~(\ref{eq:P}) are $2.1778\,{\rm N}$ and $17.6400\,{\rm N}$\,, respectively.
Therein, the first summand corresponds to the effect of the wheel contact with the surface and the second to the effect of the air resistance.

The similarity between the values obtained with expressions~(\ref{eq:P}) and (\ref{eq:power}) is a supportive evidence for the correctness of refinements provided by the latter and an indication of a reasonable accurateness of the former, in spite of its simplicity.
However, recognizing the difference between these values is consistent with the attention to marginal gains that underpin the Team GB dominance in velodrome competitions \citep[e.g.,][]{Slater2012}.

For this numerical example, ${\bf N}=[-857.4810\,,919.5203]$\,, ${\bf F}_g=[0\,,-1088.91]$\, ${\bf F}_{cp}=[-675.8359\,,0]$ and ${\bf F}_f=[181.6451\,,169.3897]$\,.
These forces are illustrated in Figure~\ref{fig:FigFreeBody}; the positive directions are upward and rightward.
As required, the resultant of the vertical forces is zero, and the resultant of the horizontal forces is equivalent to ${\bf F}_{cp}$\,.
Also, the orientation of ${\bf F}_f$ is $\theta=0.7505\,{\rm rad}\equiv43^\circ$\,, which means that it is parallel to the velodrome surface, as required.

If we consider the same laptime, namely, $t_\circlearrowleft=250/12=20.8(3)\,{\rm s}$\,, and, using expression~(\ref{eq:LapTime1}), we arbitrarily choose $v_\rightarrow = 11.7364\,{\rm m/s}\equiv \overline V_{\!\rightarrow}$ and $v_{\!\curvearrowleft}= 12.2\,{\rm m/s}$\,, which is consistent with empirical examinations of the difference between the black-line speed along the straights and along the curves, we obtain $\vartheta= 0.5701\,{\rm rad}\equiv32.6643^\circ$ and $\overline V_{\!\curvearrowleft}= 11.8565\,{\rm m/s}$\,, which results in $\overline{P}=229.2440\,{\rm W}<229.6723\,{\rm W}$\,.
This result is consistent with the fact that the same average power results in a higher black-line speed along the curves than along the straights due to the shorter distance covered by the centre of mass.
It also implies that\,---\,to achieve a given laptime\,---\,maintaining a constant black-line speed, which is tantamount to a constant cadence, does not minimize the required average power.
Nor does it imply that the minimum average power results from maintaining a constant instantaneous power.
In other words, a minimization of the average power to achieve a given laptime might not be tantamount to maintaining either a constant black-line speed or a constant power.
\subsection{Inverse problem}
For this numerical example, expression~(\ref{eq:power}) can be written as
\begin{equation}
\label{eq:Polynomial1}
\overline P=
229.6723\,{\rm W}=
993.2258\,\underbrace{\dfrac{{\rm C_{d}A}}{1-\lambda}}_X+11530.0934\,\underbrace{\dfrac{{\rm C_{rr}}}{1-\lambda}}_Y+1124.4934\,\underbrace{\dfrac{{\rm C_{sr}}}{1-\lambda}}_Z\,.
\end{equation}
In  contrast to expression~(\ref{eq:PV}), for expression~(\ref{eq:P}), and for its extension, given by expression~(\ref{eq:power}), the resistance coefficients appear only as ratios.
Hence, even with many independent equations, we cannot obtain\,---\,as an inverse solution\,---\,the values of $\rm C_{d}A$\,, $\rm C_{rr}$\,, $\rm C_{sr}$ and $\lambda$\,, but only the ratios, $X$\,, $Y$ and~$Z$\,.

To obtain the values of $X$\,, $Y$ and $Z$, we perform the least-squares fit of ten equations analogous to equation~(\ref{eq:Polynomial1}), with $\overline V_{\!\rightarrow}\in(11.5,12.5)$\,, whose matrix representation is
\begin{equation}
	\label{eq:XYZ}
\left[
		\begin{array}{c}
			250.2814 \\
			221.2244 \\
			246.4302 \\
			207.1293 \\
			210.2785 \\
			234.5504 \\
			231.1933 \\
			225.5622 \\
			221.3932 \\
			235.9688
		\end{array}
	\right]
	=
\left[
		\begin{array}{ccc}
			1090.8398 & 12040.9534 & 1008.6335 \\
			953.2952 & 11315.5242 & 1169.9259 \\
			1072.5787 & 11946.7665 & 1030.7921 \\
			886.7918 & 10950.0690 & 1242.7456 \\
			901.6366 & 11032.5717 & 1226.8169 \\
			1016.3057 & 11652.5725 & 1097.6921 \\
			1000.4207 & 11568.3931 & 1116.1798 \\
			973.7925 & 11426.0999 & 1146.7564 \\
			954.0928 & 11319.8441 & 1169.0305 \\
			1023.0198 & 11687.9987 & 1089.8239
		\end{array}
	\right]
	\left[
		\begin{array}{c}
			X \\ Y \\ Z
		\end{array}
	\right]\,.
\end{equation}
The least-squares solution of system~(\ref{eq:XYZ}) is $X=0.204082$\,, $Y=0.002041$\,, $Z=0.003061$\,.
Since $\lambda=0.02$\,, we obtain ${\rm C_{d}A}=0.2\,{\rm m^2}$\,, ${\rm C_{rr}}=0.002$\,, ${\rm C_{sr}}=0.003$\,, as expected.

For the measured values of $P$\,---\,as opposed to the modelled ones, for which three equations suffice to solve for $X$\,, $Y$ and $Z$\,---\,a redundancy of the laptime information allows us to estimate them, and to obtain statistical information about the empirical adequacy of a model, which, however sophisticated, remains only a mathematical analogy for a physical realm.
This redundancy could correspond to different laptimes during a single ride.

An insight into the consistency of information can be gained by writing each equation of system~(\ref{eq:XYZ}) as
\begin{equation*}
1=
\dfrac{a}{\overline P}\,\underbrace{\dfrac{{\rm C_{d}A}}{1-\lambda}}_X+\dfrac{b}{\overline P}\,\underbrace{\dfrac{{\rm C_{rr}}}{1-\lambda}}_Y+\dfrac{c}{\overline P}\,\underbrace{\dfrac{{\rm C_{sr}}}{1-\lambda}}_Z\,,
\end{equation*}
and plotting $a/\overline P$\,, $b/\overline P$ and $c/\overline P$\,.
For system~(\ref{eq:XYZ}), they are collinear.
For measurements, the departure from the collinearity is indicative of the quality of the model and of the measurement errors.
Within a model, such a plot can be used to study the sensitivity of $X$\,, $Y$ and~$Z$ to perturbations.

To estimate $\rm C_{d}A$\,, $\rm C_{rr}$ and $\rm C_{sr}$\,, the value of $\lambda$ needs to be given independently or be assumed.
It is commonly accepted that, for high-quality track bicycles, $\lambda\in(0.01,0.03)$\,.
Also, if the power meter is in the rear hub, as opposed to being in the pedals or the bottom bracket, $\lambda\approx 0$\,, since the effect of the resistance of the drivetrain\,---\,which includes the chainring, chain and sprocket\,---\,upon the measuring device is nearly eliminated \citep[e.g.,][]{Chung2012}.

In general, given the laptime\,---\,and assuming $v_\rightarrow$ to be known\,---\,we can use equation~(\ref{eq:LapTime}) to find $\vartheta$\,, which in turn can be substituted into equation~(\ref{eq:Num3}) to find $\overline V_{\!\curvearrowleft}$\,, which can then be substituted into equation~(\ref{eq:LapTime2}) to find $\overline v_{\!\curvearrowleft}$\,.
In other words, given the laptime and the speed along the straights, the lean angle is implicitly determined, and in turn determines the centre-of-mass and black-line speeds along the curves.
\section{Conclusions}
Expression~(\ref{eq:power}), together with expressions~(\ref{eq:LapTime1}), (\ref{eq:LapTime2}) and (\ref{eq:Num3}), allows us to calculate the power required to achieve a desired individual-pursuit time or the time achievable with a particular power.
It also allows us to quantify the effects of the bicycle-cyclist weight, air resistance, rolling resistance, drivetrain resistance and lateral friction, as well as of the velodrome size, steepness of its banks and tightness of its curves.
Furthermore, a quantification of these effects lends itself to a study of optimization of a cyclist's effort \citep[][Appendix~A]{DSSBici2}.

Our formulation and examination of an inverse solution for expression~(\ref{eq:power}) shows that we can infer only the ratios, ${\rm C_dA}/(1-\lambda)$\,, ${\rm C_{rr}}/(1-\lambda)$ and ${\rm C_{sr}}/(1-\lambda)$\,.
Nevertheless, the proposed solution allows us to gain an insight into the consistency between the measurements and the model.

As illustrated by the presented results, accurate inferences based on the power-meter measurements on a velodrome require a distinction between the trajectory of wheels, which we assume to coincide with the black line, and the trajectory of the centre of mass.
The forces involved and relations among them, as well as other entailed quantities, are assumed to be functions of the latter.

If necessary, to model highly accurate measurements or for a specific scope of investigation, the mathematical model stated in expression~(\ref{eq:power}) can be refined in a manner suggested in footnotes~\ref{foot:Straight} and~\ref{foot:Wheels}.
In contrast to these refinements, a progressive leaning and straightening\,---\,neglected to formulate the two distances, in expression~(\ref{eq:DistLap}), and resulting in two constant centre-of-mass speeds,~$\overline V_{\!\rightarrow}$ and $\overline V_{\!\curvearrowleft}$\,---\,cannot be achieved within this model, as discussed in Appendix~\ref{app:Trans}; it is a subject of the subsequent article \citep{BSSSbici4}.

The results presented in this article might be an {\it a posteriori} reassurance and comfort for Michael~\citet[p.~251]{Hutchinson}, in his attempt to achieve immortality the hard way,
\begin{quote}
Ride fast---in the end that's all it ever comes down to.
The pressure of another curve, the relief of the simple straight.
But the straight's short respite is never enough.
My shoulders are awful.
My arms hurt.
And every few seconds I have to manage another banked curve.
Each one demands a little more effort, a little more concentration.
As the physical toll mounts, the balance and rhythm aren't offering the protection from reality that they did.
I need some sort of reassurance.
Some comfort.	
\end{quote}
The power required to maintain the same speed\,---\,with respect to the black line\,---\,is less on a curve than on a straight.
In a certain manner, the curves provide a short respite, since\,---\,thanks to the curves\,---\,the distance travelled by a centre of mass, within a given time, is shorter than the distance with respect to the black line.
\begin{appendix}
\section{Lean angle}
\label{app:LeanAng}
\setcounter{equation}{0}
\setcounter{figure}{0}
\renewcommand{\theequation}{\Alph{section}.\arabic{equation}}
\renewcommand{\thefigure}{\Alph{section}\arabic{figure}}
Expressions~\eqref{eq:vartheta} and \eqref{eq:Fc_vartheta} imply that the lean angle,~$\vartheta$\,, depends only on the centripetal acceleration of the cyclist's centre of mass and the acceleration of gravity.
In other words, the lean angle depends only on the centre-of-mass speed and the radius of curvature of the centre-of-mass trajectory, not on the track inclination,~$\theta$\,, even though both the normal force, stated in expression~\eqref{eq:N}, and the frictional force,  in expression~\eqref{eq:Ff}, do depend on track inclination.
However, the $\theta$-dependence cancels out of the centripetal force, stated in expression~\eqref{eq:Fc_vartheta}, by a seldom used trigonometric identity.

Given the generality of this result, it would be satisfying to obtain it in a manner that explains it in the context of physics.
To this end, let us analyze the situation from inside the noninertial frame of the cyclist.
Specifically, we consider the frame comoving with the cyclist around a curve, with the cyclist considered as a point mass.
We neglect the additional accelerated motion resulting from the rotation of the cyclist about an axis through the centre of mass.

\begin{figure}[h]
\centering
\includegraphics[scale=0.7]{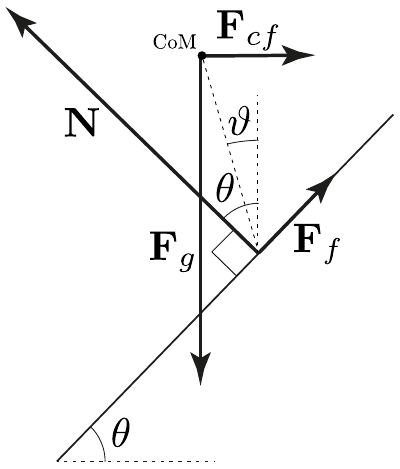}
\caption{\small Force diagram: Noninertial frame}
\label{fig:FigNonIner}
\end{figure}

As illustrated in Figure~\ref{fig:FigNonIner} and in contrast to Figure~\ref{fig:FigFreeBody}, in this noninertial frame, instead of the forces in the horizontal direction summing to the centripetal force,~${\bf F}_{cp}$\,, the forces in this direction, including the fictitious centrifugal force, ${\bf F}_{cf}=-{\bf F}_{cp}$\,, sum to zero.
The centrifugal force must be taken to act at the centre of mass, since otherwise the torque about the centre of mass, stated in expression~\eqref{eq:torque=0}, would be affected.

To proceed, we invoke the vector identity,
\begin{equation*}
    \sum{\boldsymbol\tau} = {\bf R}_{\rm\scriptscriptstyle CoM}\times\sum{\bf F} + \sum{\boldsymbol\tau}_{\rm\scriptscriptstyle CoM}\,,
\end{equation*}
where $\sum{\boldsymbol\tau}$ is the net torque about an arbitrary point, $\sum{\boldsymbol\tau}_{\rm\scriptscriptstyle CoM}$ is the net torque about the centre of mass, ${\bf R}_{\rm\scriptscriptstyle CoM}$ is the position vector of the centre of mass, and $\sum{\bf F}$ is the net force.
From this identity there follows the well-known result that if the net force is zero and the net torque about the centre of mass is zero, the net torque about any other point is also zero.
In particular, let us consider the torque about the point of contact of the tires with the surface,
\begin{equation*}
\sum\tau_z = h\,F_g\sin\vartheta-h\,F_{cf}\sin\left(\frac{\pi}{2}-\vartheta\right) = 0\,,
\end{equation*}
from which\,---\,considering the magnitudes\,---\,it follows that
\begin{align}
\label{eq:lean_angle}
\nonumber\tan{\vartheta} &= \frac{F_{cf}}{F_g}\\
&= \frac{F_{cp}}{F_g},
\end{align}
where expression~\eqref{eq:lean_angle} is equivalent to expressions~\eqref{eq:vartheta} and \eqref{eq:Fc_vartheta}.
Expression~\eqref{eq:lean_angle} manifestly holds whether the curve is banked or unbanked.

Physically, from inside the noninertial frame of the cyclist, the value of the lean angle that obtains for specific values of speed and radius is the one that makes the gravitational torque balance the centrifugal torque.
In particular, this condition makes no reference to track inclination.
\section{Harmonic mean}
\label{app:HarmAve}
\setcounter{equation}{0}
\setcounter{figure}{0}
\renewcommand{\theequation}{\Alph{section}.\arabic{equation}}
\renewcommand{\thefigure}{\Alph{section}\arabic{figure}}
To calculate the average power over a lap, we require total work and time,
\begin{equation*}
	W = \oint F(s)\,{\rm d}s\quad{\rm and}\quad
	T = \oint{\rm d}t = \oint\frac{{\rm d}s}{V(s)}\,,
\end{equation*}
to write
\begin{equation*}
	\overline{P} 
	= 
	\frac{W}{T} 
	=
	\dfrac{{\displaystyle\oint F(s)\,{\rm d}s}}{{\displaystyle\oint\frac{{\rm d}s}{V(s)}}}\,.
\end{equation*}
If we consider $n$ segments along which $F$ and $V$ are constant, we write 
\begin{equation*}
\overline{P} 
=
	\dfrac{{\displaystyle\sum_{i=1}^nF_{i}\,\Delta s_{i}}}{{\displaystyle\sum_{i=1}^n\dfrac{\Delta s_{i}}{V_{i}}}}\,{\displaystyle\dfrac{\dfrac{1}{\Delta s}}{\dfrac{1}{\Delta s}}}
=
	\dfrac{{\displaystyle\sum_{i=1}^nF_{i}\dfrac{\Delta s_{i}}{\Delta s}}}{{\displaystyle\sum_{i=1}^n\dfrac{1}{V_{i}}}\dfrac{\Delta s_{i}}{\Delta s}}
	\,.
\end{equation*}
Since in our case, there are only two segments, we write
\begin{equation}
\label{eq:AveP}
	\overline{P} 
	=
	\dfrac{F_{\!\rightarrow}\dfrac{\Delta s_{\rightarrow}}{\Delta s} + F_{\!\curvearrowleft}\dfrac{\Delta s_{\curvearrowleft}}{\Delta s}}{\dfrac{1}{V_{\!\rightarrow}}\dfrac{\Delta s_{\rightarrow}}{\Delta s}+\dfrac{1}{V_{\!\curvearrowleft}}\dfrac{\Delta s_{\curvearrowleft}}{\Delta s}}
	\,,
\end{equation}
where ${}_{\rightarrow}$ denotes straights and ${}_{\curvearrowleft}$ denotes banks.
The numerator is the distance-weighted arithmetic mean,~$\overline{F}$\,.
The denominator is reciprocal of the distance-weighted harmonic mean,
\begin{equation*}
	\langle V\rangle
	=
	\dfrac{\displaystyle\sum_{i=1}^2\dfrac{\Delta s_{i}}{\Delta s}}{\displaystyle\sum_{i=1}^2\dfrac{1}{V_{i}}\dfrac{\Delta s_{i}}{\Delta s}}
	=
	\frac{1}{\dfrac{1}{V_{\!\rightarrow}}\dfrac{\Delta s_{\rightarrow}}{\Delta s}+\dfrac{1}{V_{\!\curvearrowleft}}\dfrac{\Delta s_{\curvearrowleft}}{\Delta s}}\,,
\end{equation*}
which\,---\,for the average centre-of-mass speed\,---\,results in expression~(\ref{eq:AveVa}).

In contrast to $\langle V\rangle$\,, the distance-weighted arithmetic mean is
\begin{equation}
\label{eq:Vmean}
	\overline{V}
	=
	\dfrac{\displaystyle\sum_{i=1}^2V_{i}\,\Delta s_{i}}{\Delta s}
	=
	{V_{\!\rightarrow}}\dfrac{\Delta s_{\rightarrow}}{\Delta s}+V_{\!\curvearrowleft}\dfrac{\Delta s_{\curvearrowleft}}{\Delta s}\,,
\end{equation}
which, in view of expression~(\ref{eq:AveP}), confirms that $\overline{P}\neq\overline{F}\,\overline{V}$\,.
The harmonic mean is less than the arithmetic mean; it is skewed toward slower speeds.
Thus, $\overline{F}\,\overline{V}$ would overestimate the average power.

For our numerical example, following expression~(\ref{eq:AveV}), $\langle{V}\rangle=11.8091\,{\rm m/s}$\,; hence, in accordance with expression~(\ref{eq:power}), $\overline{F}\,\langle{V}\rangle=229.6723\,{\rm W}$\,.
On the other hand, following expression~(\ref{eq:Vmean}), $\overline{V}=11.8114\,{\rm m/s}$\,, and $\overline{F}\,\overline{V}=229.7161\,{\rm W}$\,.
The difference is small due to the similarity of values of $\overline V_{\!\rightarrow}$ and $\overline V_{\!\curvearrowleft}$\,.
It would not be so, for an average of an upwind and downwind segments, discussed by \citet[Appendix~A]{DSSBici2}, where also the harmonic mean is used.
\section{Transition between curves and straights}
\label{app:Trans}
\setcounter{equation}{0}
\setcounter{figure}{0}
\renewcommand{\theequation}{\Alph{section}.\arabic{equation}}
\renewcommand{\thefigure}{\Alph{section}\arabic{figure}}
In this article, to consider the velodrome in question, we assume that, along the banks, the radius of curvature is constant.
Hence, the track is composed of two semicircles and two straights; this is the case of the light grey oval in Figure~\ref{fig:FigVelo}.
The dark grey and black ovals also represent a track whose $S=250\,{\rm m}$\,, but their radii of curvature are not constant; they are $r=23\,{\rm m}$\,, at the intercepts with the horizontal axis, and $r\to\infty$\,, at the intercepts with the vertical axis.
The dark grey oval is composed of an ellipse \citep{Brachistochrone}, whose semiaxes are $23$ and $30$\,, and of two straights.
The black oval, in polar coordinates, is
\begin{equation}
\label{eq:rphi}
r(\phi)=28.15\left(1+0.5\,\sin^2\!\!\phi\,\cos^2\!\!\phi+0.7\,\cos^4\!\!\phi\right)\,,\qquad\phi=[\,0,2\pi)\,,
\end{equation}
where the coefficients are found numerically by invoking the concept of the arclength and curvature to ensure that $S=250\,{\rm m}$\,, $r_{\rm min}=23\,{\rm m}$\,, $r_{\rm max}\to\infty$\,.
\begin{figure}[h]
\centering
\includegraphics[scale=0.7]{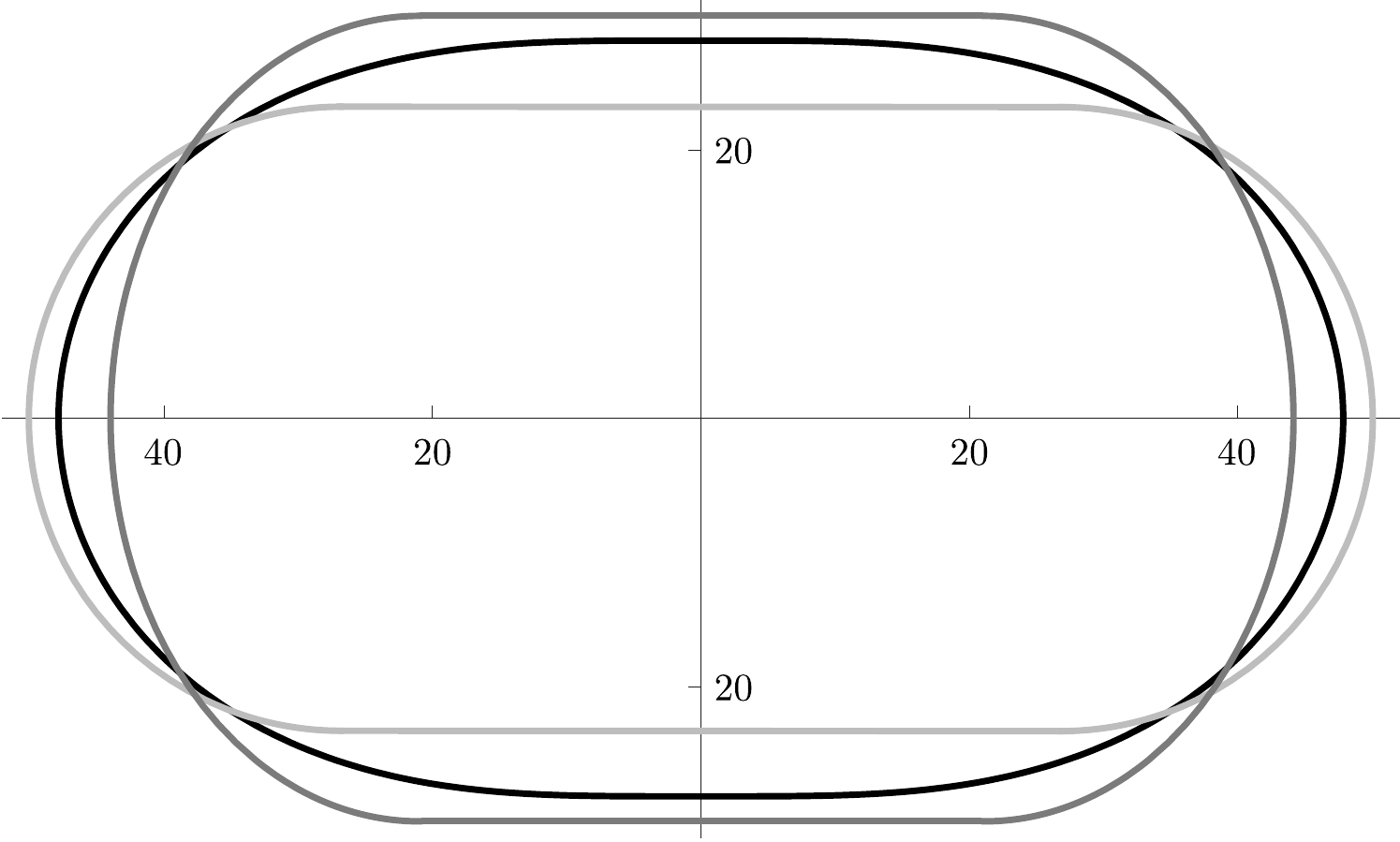}
\caption{\small Velodrome tracks}
\label{fig:FigVelo}
\end{figure}

These ovals share important geometrical properties, namely, their circumference and their radii of curvature, at the horizontal-axis and vertical-axis intercepts.
However, the model presented in this article applies explicitly to the light grey oval.
Its application to other similar ovals entails a decrease in accuracy.

For the light-grey oval, a model requires two constant centre-of-mass speeds, for the straights and for the circular banks.
For the dark-grey oval, an explicit model requires the centre-of-mass speed to be represented by two functions, where the speed along the straight is constant, but along the banks is not, due to the changing radius of curvature along the elliptical bank.
For the black oval, an explicit model requires the centre-of-mass speed to be represented by a single function, which depends on the continuously changing radius of curvature. 

\begin{figure}[h]
\centering
\includegraphics[scale=0.7]{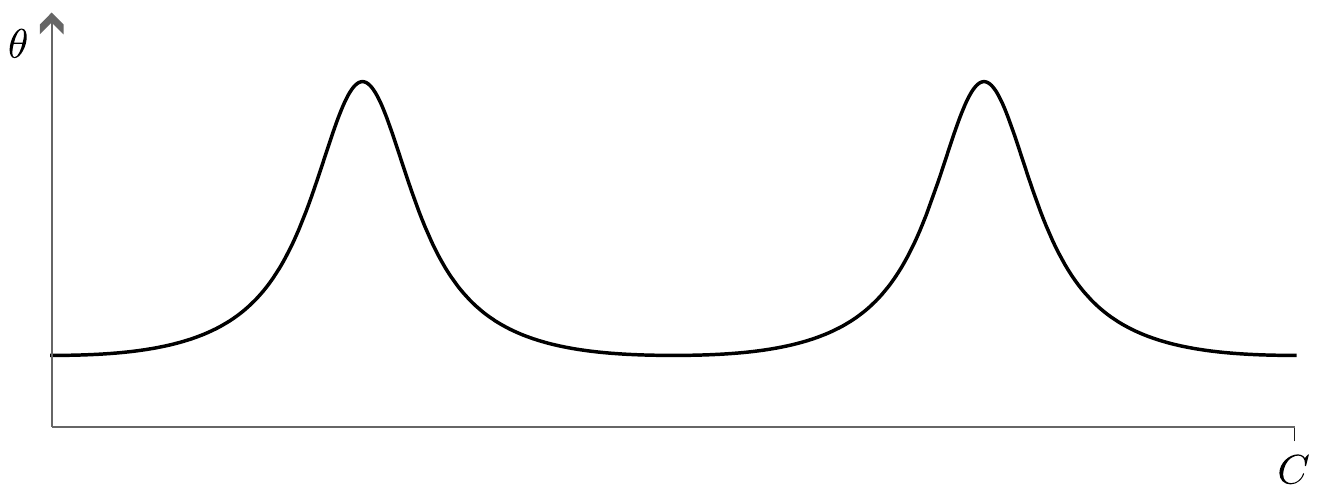}
\caption{\small Track inclination}
\label{fig:FigSlope}
\end{figure}

Let us examine the black-oval model.
Its average curvature, for the length of a lap, is
\begin{align*}
\overline\kappa=\dfrac{\displaystyle\int\limits_0^S\kappa(s)\,{\rm d}s}{\displaystyle\int\limits_0^S{\rm d}s}
&=\dfrac{\displaystyle\int\limits_0^{2\pi}\dfrac{\left|r^2(\phi)+2\left(\dfrac{\partial r(\phi)}{\partial\phi}\right)^2-r(\phi)\,\dfrac{\partial^2r(\phi)}{\partial\phi^2}\right|}{\left(\sqrt{r^2(\phi)+\left(\dfrac{\partial r(\phi)}{\partial\phi}\right)^2}\right)^3}\sqrt{r^2(\phi)+\left(\dfrac{\partial r(\phi)}{\partial\phi}\right)^2}\,{\rm d}\phi}
{\displaystyle\int\limits_0^{2\pi}\sqrt{r^2(\phi)+\left(\dfrac{\partial r(\phi)}{\partial\phi}\right)^2}\,{\rm d}\phi}\\\\
&=\dfrac{\displaystyle\int\limits_0^{2\pi}\dfrac{\left|r^2(\phi)+2\left(\dfrac{\partial r(\phi)}{\partial\phi}\right)^2-r(\phi)\,\dfrac{\partial^2r(\phi)}{\partial\phi^2}\right|}{r^2(\phi)+\left(\dfrac{\partial r(\phi)}{\partial\phi}\right)^2}\,{\rm d}\phi}
{\displaystyle\int\limits_0^{2\pi}\sqrt{r^2(\phi)+\left(\dfrac{\partial r(\phi)}{\partial\phi}\right)^2}\,{\rm d}\phi}\,;	
\end{align*}
a numerical integration, with $r$ given in expression~(\ref{eq:rphi}), results in $\overline\kappa=0.0251257\,{\rm m^{-1}}$\,.
Hence, the average radius of curvature is $\overline r:=1/\,\overline\kappa=39.7998\,{\rm m}$\,.
This is consistent with an expectation in view of the light-grey oval, whose radius of curvature is $r=23\,{\rm m}$\,, along the banks, and infinity along the straights.
For the black-oval model, in a manner analogous to expressions~(\ref{eq:radii}) and (\ref{eq:DistLap}), the distance traveled\,---\,in one lap\,---\,by the centre of mass is
\begin{align*}
&\int\limits_0^{2\pi}\sqrt{\Big(r(\phi)-h\sin\overline\vartheta
\Big)^2+\left(\dfrac{\partial\Big(r(\phi)-h\sin\overline\vartheta
\Big)}{\partial\phi}\right)^2}{\rm d}\phi\\
=&\int\limits_0^{2\pi}\sqrt{\Big(r(\phi)-h\sin\overline\vartheta
\Big)^2+\left(\dfrac{\partial r(\phi)}{\partial\phi}\right)^2}{\rm d}\phi\,,
\end{align*}
where $\overline\vartheta$ is the average lean angle.
Thus, given a laptime, in a manner analogous to expression~(\ref{eq:LapTime2}), we write
\begin{equation}
\label{eq:tblack}
t_\circlearrowleft=\dfrac{\displaystyle\int\limits_0^{2\pi}\sqrt{\Big(r(\phi)-h\sin\overline\vartheta
\Big)^2+\left(\dfrac{\partial r(\phi)}{\partial\phi}\right)^2}{\rm d}\phi}{\overline V}\,,
\end{equation}
where $\overline V$ is the average speed, which\,---\,following expression~(\ref{eq:Num3})\,---\,we write as
\begin{equation}
\label{eq:Vblack}
\overline V=\sqrt{g\,\overline r\,\tan\vartheta}\,.
\end{equation}
\begin{figure}[h]
\centering
\includegraphics[scale=0.7]{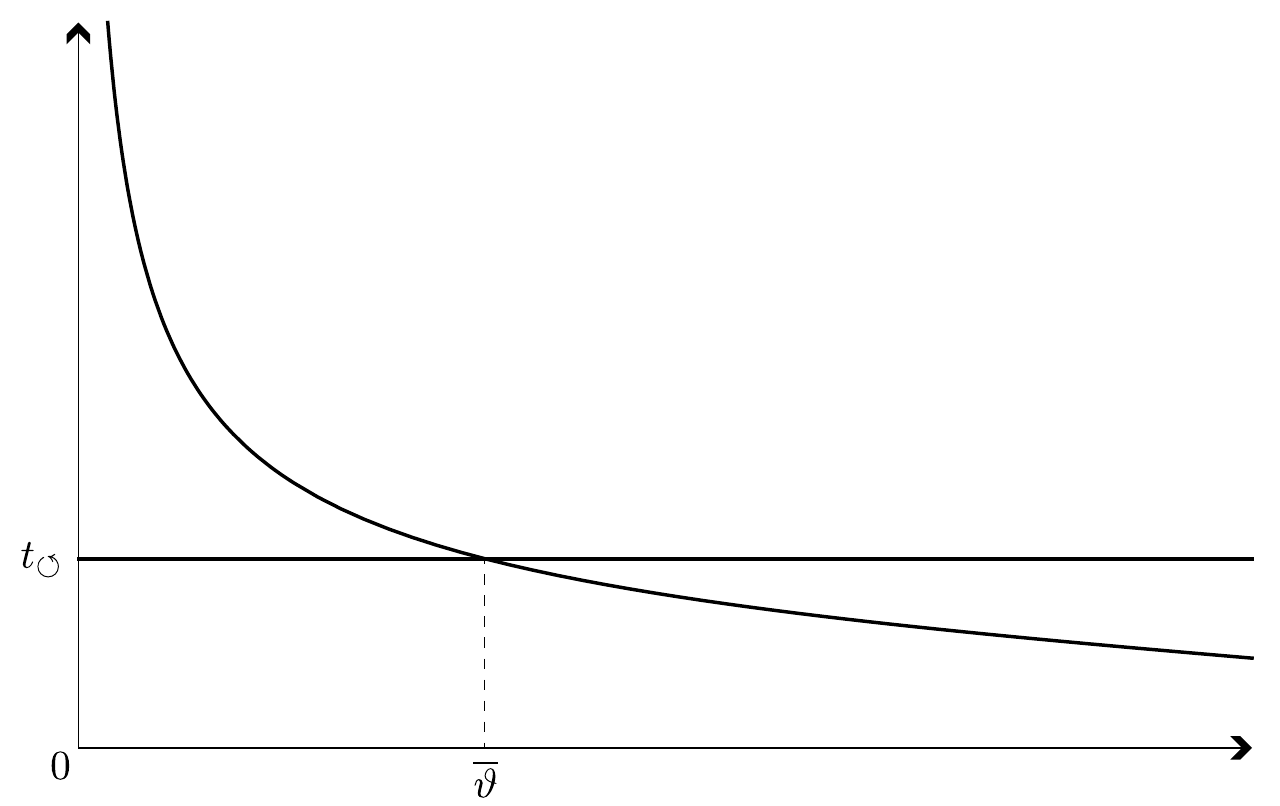}
\caption{\small Left-hand and right-hand sides of equation~(\ref{eq:tblack})}
\label{fig:FigSol}
\end{figure}
To compare the resulting power with the numerical example in Section~\ref{sec:NumEx}, we let the laptime be such that, according to expression~(\ref{eq:LapTime1}), the corresponding black-line speed is~$\overline V_{\!\rightarrow}=12\,{\rm m/s}$\,.
In accordance with expressions~(\ref{eq:tblack}) and (\ref{eq:Vblack}), and as shown in Figure~\ref{fig:FigSol}, we obtain $\overline\vartheta=0.347284\,{\rm rad}\equiv 19.8979^\circ$\,, numerically, which results in $\overline V=11.8878\,{\rm m/s}$\,.
Hence, in accordance with expression~(\ref{eq:P}),~$\overline P=236.415\,{\rm W}$\,.%
\footnote{We could refine the black-oval model by including the effect of the track inclination, which\,---\,given a minimum and maximum values of inclination along the oval, stated in expression~(\ref{eq:rphi}), as well as an interpolation formula between them\,---\,is illustrated in Figure~\ref{fig:FigSlope}.
Hence, this effect could be expressed as a continuous function of distance.}
 If we consider the average centre-of-mass speed, per lap, for the light-grey oval, which\,---\,in accordance with expression~(\ref{eq:AveV})\,---\,is $\overline V=11.8114\,{\rm m/s}$\,, we obtain $\overline P=232.223\,{\rm W}$\,.
Another comparison is the distance travelled by the centre of mass.
For the black oval, it is~$247.662\,{\rm m}$\,; for the light-grey oval, it is~$246.024\,{\rm m}$\,.
 
The model based on the light-gray oval requires fewer approximations, within the realm of mathematics.
This is a consequence of the idealization of a velodrome track, which is greater for the light-gray oval than for the black oval, due to the assumption of a constant curvature and no transition between the curves and the straights.
 Thus, in spite of more mathematical approximations, the latter might exhibit a superior empirical adequacy.
A~further examination of this question, which is essential to the concept of modelling, requires experimental results.
The conclusiveness of such results, however, might also be questionable, in view of the similarity of $\overline P=236.35\,{\rm W}$\,, for the black oval, and $\overline P=229.7161\,{\rm W}$\,, for the light-grey oval.
\end{appendix}
\section*{Acknowledgements}
We wish to acknowledge Len Bos for his mathematical insights, Roger Mason for his perceptive comments, David Dalton, for his scientific editing and proofreading, Elena Patarini, for her graphic support, and Roberto Lauciello, for his artistic contribution. 
Furthermore, we wish to acknowledge Favero Electronics for inspiring this study by their technological advances and for supporting this work by providing us with their latest model of Assioma Duo power meters.
\section*{Conflict of Interest}
The authors declare that they have no conflict of interest.
\bibliographystyle{spbasic}
\bibliography{SSSbici3_arXiv.bib}
\end{document}